\documentclass[11pt]{article}
\usepackage{moriond,epsfig}

\def\NIMA{{\em Nucl. Instrum. Methods} A}

\def\PLB{{\em Phys. Lett.}  B}

\def\be{\begin{equation}}
\def\ee{\end{equation}}
\def\bea{\begin{eqnarray}}
\def\eea{\end{eqnarray}}

\begin{document}
\vspace*{4cm}
\title{SEARCH FOR SQUARKS AND GLUINOS AT D\O}

\author{ S. CALVET}

\address{Centre de Physique des Particules de Marseille,\\
 IN2P3-CNRS, Universit\'e de la M\'editerran\'ee, \\
163, Av. de Luminy, Marseille, France }

\maketitle\abstracts{A search for scalar quarks and gluinos is performed with 0.96~fb$^{-1}$ of data collected by the D\O~experiment in $p\bar{p}$ collisions at $\sqrt{s}=$1.96~TeV at the Fermilab Tevatron Collider. The topologies analyzed consist of acoplanar jets and  multijet events with missing transverse energy. We find the data to be consistent with Standard Model expectations, and set 95\% C.L. exclusion domains in the squark and gluino mass plane and in the ($m_0$, $m_{1/2}$) plane, within the framework of minimal supergravity with tan$\beta=3$, $A_0=0$ and $\mu<0$.}

\def\Missing#1#2{{\mbox{$#1\kern-0.57em\raise0.19ex\hbox{/}_{#2}$}}\ }
\def\met{\mbox{$\Missing{E}{T}$}}
\def\dijet{{\bf dijet~}}
\def\3jet{{\bf 3-jets}}
\def\multi{{\bf gluino~}}
\def\Ht{$H_T$~}

\section{Introduction}
Supersymmetric theories predict for each elementary fermion (resp. boson) a new elementary boson (resp. fermion). These new particles have the same quantum numbers as their Standard Model (SM) partners, but a spin which differs by one half. Particles carry a new multiplicative quantum number $R$, which is  $1$ for SM particles and $-1$ for super-particles.  The partners of quarks (resp. gluons) are called squarks (resp. gluinos) and have a spin $0$ (resp. $1/2$). 

In R-parity conserving theories, supersymmetric particles are thus produced in pairs and final states of their decay chains contain the lightest supersymmetric particles (LSP) which is stable. The supersymmetric partners of neutral gauge and Higgs bosons are called neutralinos, and the lightest one, $\tilde{\chi}_1^0$, arises as the natural LSP in supergravity inspired models\cite{cite:mSUGRA}.

\begin{figure}[t] 
\caption{Main decays of squark ({\bf a}) and gluino ({\bf b}).
\label{fig:DiagFeynman}}
\begin{center}
\begin{minipage}{0.18\linewidth}
	\centering
	\psfig{figure=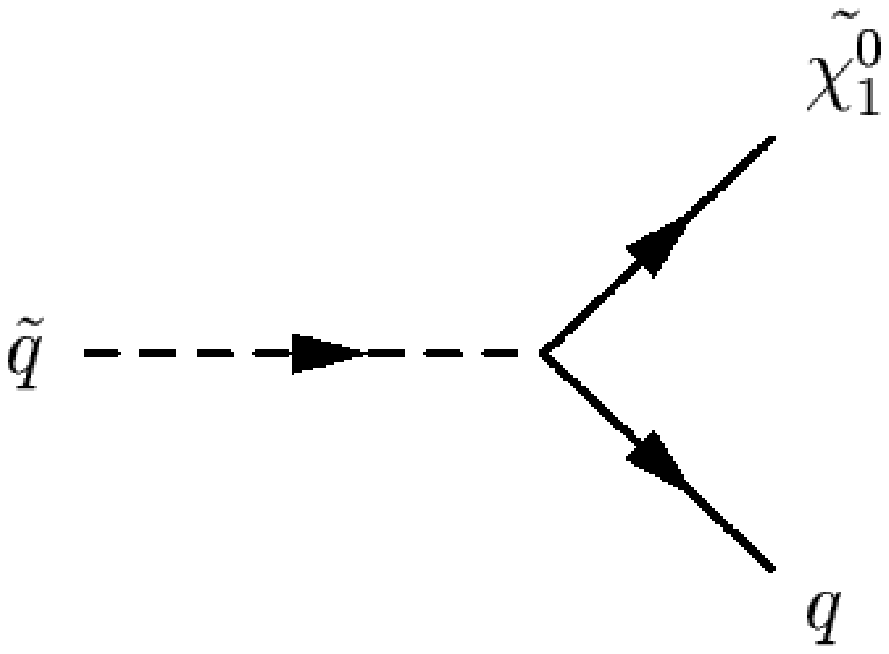,width=\linewidth} \\ \vspace{-0.7cm}(a)
\end{minipage}
\begin{minipage}{0.28\linewidth}
	\centering
	\psfig{figure=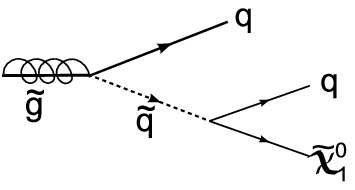,width=\linewidth} \\ \vspace{-1.cm}(b)
\end{minipage}
\end{center}
\end{figure}

The generic models predict squarks of the four lightest quark flavors to have similar masses, and this, independently of the helicity of the quark. If they are sufficiently light, squarks can be widely produced at Tevatron and they decay according to $\tilde{q}\to q\tilde{\chi}_1^0$ as shown in Fig.~\ref{fig:DiagFeynman}.a. Tevatron can also produce a large amount of gluinos if they are not too heavy. In particular, if they are lighter than squarks, their preferred decay is then $\tilde{g}\to q\bar{q}\tilde{\chi}_1^0$ (Fig.~\ref{fig:DiagFeynman}.b).

\begin{figure}[t] 
\caption{Main squarks and gluinos production diagrams. Diagram {\bf a} (resp. {\bf c}) dominates if squarks (resp. gluinos) are lighter than gluinos (resp. squarks). If squarks and gluinos have similar masses, contribution of diagram {\bf b} is large.
\label{fig:DiagFeynmanProd}}
\begin{center}
\begin{minipage}{0.19\linewidth}
	\centering
	\psfig{figure=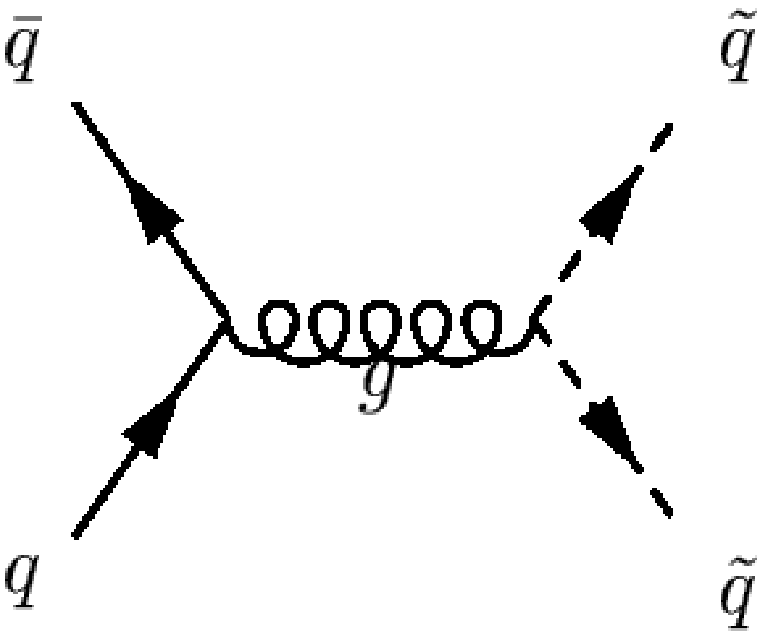,width=\linewidth} \\ \vspace{-0.5cm} (a)
\end{minipage}
\begin{minipage}{0.20\linewidth}
	\centering
	\psfig{figure=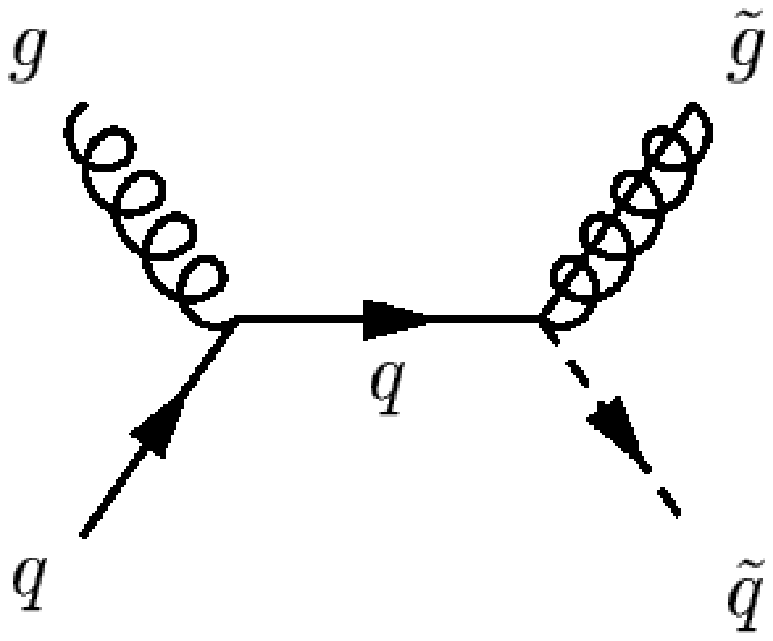,width=\linewidth} \\ \vspace{-0.5cm} (b)
\end{minipage}
\begin{minipage}{0.20\linewidth}
	\centering
	\psfig{figure=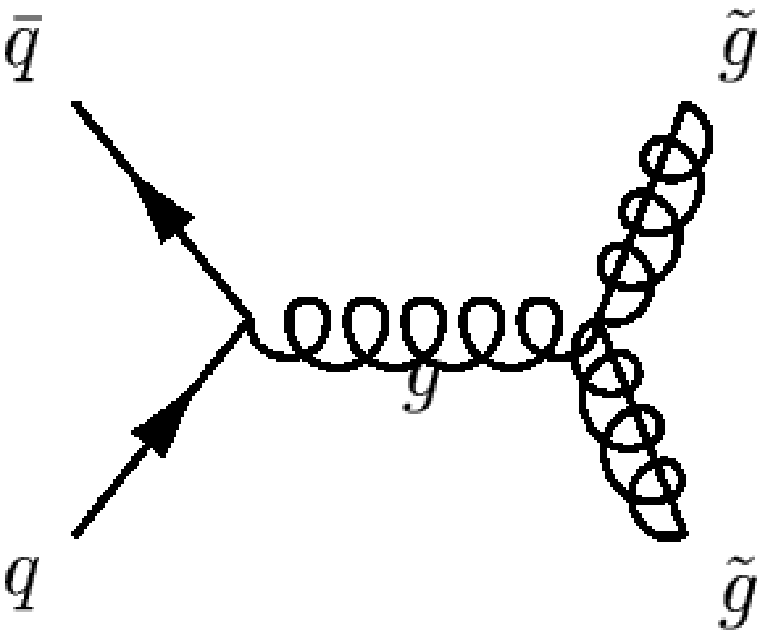,width=\linewidth} \\ \vspace{-0.5cm} (c)
\end{minipage}
\end{center}
\end{figure}

The $\tilde{\chi}_1^0$ not being detected, final state topologies for such events are two jets and missing transverse energy (\met) in the case of squark pair production (Fig.~\ref{fig:DiagFeynmanProd}.a), at least three jets and \met for simultaneous production of squark and gluino (Fig.~\ref{fig:DiagFeynmanProd}.b), and four jets and \met for pair production of gluinos (Fig.~\ref{fig:DiagFeynmanProd}.c).

This search is performed in the minimal supergravity (mSUGRA) framework\cite{cite:mSUGRA}, with the following parameters fixed: $tan\beta=3$, $A_0=0$ and $\mu<0$. In order to increase the sensitivity, three analyses are developed, searching events with two (\dijet analysis), three (\3jet analysis) or at least four (\multi analysis) jets and \met.

\section{Common part of the three analyses}
The three analyses start with a common data set and preselection which are later on optimized. The data sample used corresponds to an integrated luminosity of 0.96~fb$^{-1}$. At the trigger level it selects events with 2 acoplanar jets and \met, or with multi-jets and \met.

\begin{figure}[t] 
\caption{{\bf (a)} Distribution in \met before the two final cuts on $H_T$ and \met in the \dijet analysis. The signal $M_{\tilde{q}}=375$~GeV and $M_{\tilde{g}}=416$~GeV is drawn as a hatched histogram. 
{\bf (b)} Distribution in $\Delta\phi_{min}($\met, any jet$)$ in the \dijet analysis before cutting on that variable. 
\label{fig:MET_Slection}}
\begin{center}
\begin{minipage}{0.40\linewidth}
\begin{center}
\psfig{figure=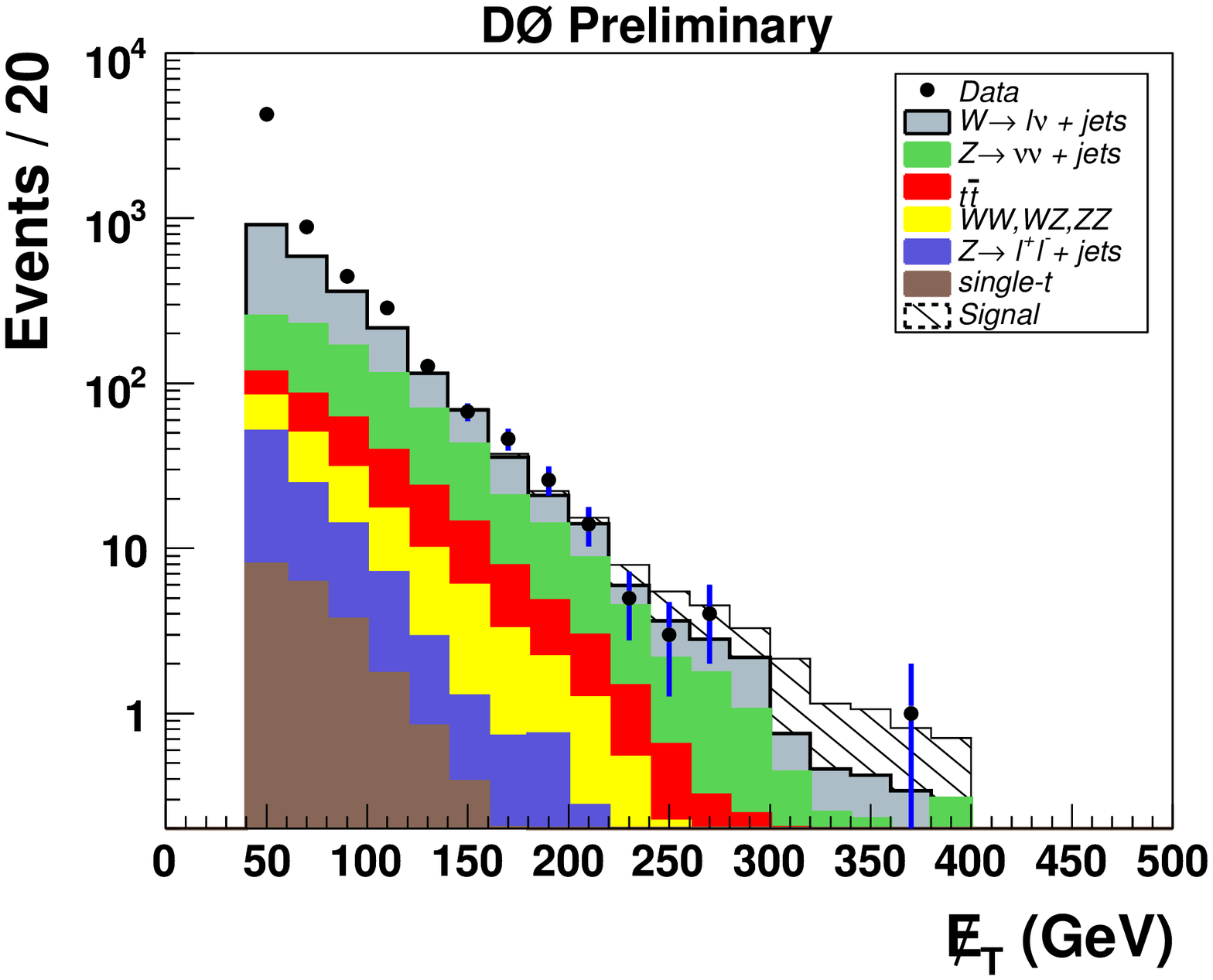,width=\linewidth}\\ 
\end{center}
\vspace{-0.6cm}(a) 
\end{minipage}
\begin{minipage}{0.4\linewidth}
\begin{center}
\psfig{figure=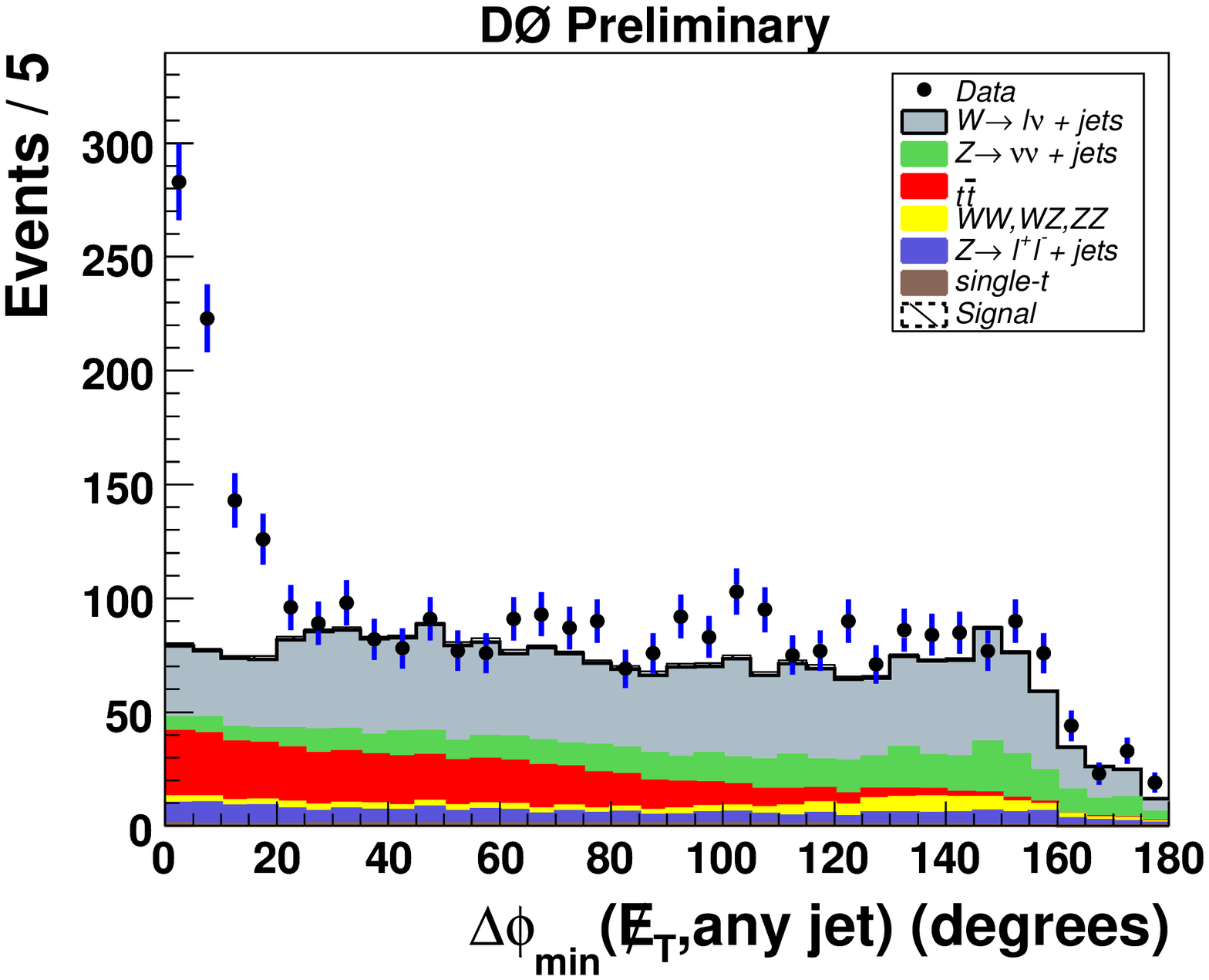,width=\linewidth}\\
\end{center}
\vspace{-0.6cm}(b)
\end{minipage}
\end{center}
\end{figure}

The main backgrounds are multijet events with \met coming from mismeasured jets(instrumental background), and "electroweak" background: $Z^0(\nu\nu)+jets$, $W^\pm(l\nu)+jets$ (with $l$ a charged lepton) and $t\bar{t}$. Because the signals have jets with high transverse energy ($E_T$), one selects events with at least two jets with $E_T>35$~GeV in central region of the calorimeter ($|\eta|<0.8$, where $\eta=ln[tan(\theta/2)]$, and $\theta$ is the polar angle relative to proton beam). In order to reduce instrumental background, jets are track confirmed and the two leading jets must be acoplanar: $\Delta\phi(jet_1, jet_2)<165^\circ$ ($\phi$ being the azimuthal angle). The instrumental background is greatly reduced by requiring \met$>$75~GeV (Fig.~\ref{fig:MET_Slection}.a) and an isolation of \met. For example, the minimal angle between \met and jets has to be greater than 40$^\circ$ for the \dijet analysis (Fig.~\ref{fig:MET_Slection}.b). Finally veto on isolated muon and electron is performed to fight against electroweak background.

\section{Optimization and combination of the three analyses}
After preselection the three analyses are optimized by adjusting cuts on \met and \Ht which is the scalar sum of the $E_T$ of the jets. The optimized values of \met and \Ht and listed in Tab.~\ref{tab:OptiCut}.a. No excess of events is found, data being in agreement with SM expectation for all analyses (Tab.~\ref{tab:OptiCut}.b).
%
\begin{table}
\caption{Results of the optimization.\label{tab:OptiCut}}
\centering
\begin{tabular}{|l|c|c|  p{0.15in}  |l|c|ccc|}
\multicolumn{3}{c}{\footnotesize{Table~\ref{tab:OptiCut}.a: Optimized cuts (GeV).}} & \multicolumn{1}{c}{}  & 
\multicolumn{5}{c}{\footnotesize{Table~\ref{tab:OptiCut}.b: Remaining events for optimized analyses.}}\\
\cline{1-3}\cline{5-9}
analysis &\mbox{~~~}\met\mbox{~~~}& \Ht & & analysis & data & \multicolumn{3}{c|}{total background} \\
\cline{1-3}\cline{5-9}
\dijet   &  225  & 300 & & \dijet   &   5  &   7.5      & $\pm 1.1$ (stat.) & $^{+1.3}_{-1.0}$ (syst.)\\
\3jet    &  150  & 400 & & \3jet    &   6  &   6.1      & $\pm 0.4$ (stat.) & $^{+1.3}_{-1.2}$ (syst.)\\
\multi   &  100  & 300 & & \multi   &  34  &  33.4      & $\pm 0.8$ (stat.) & $^{+5.6}_{-4.9}$ (syst.)\\
\cline{1-3}\cline{1-3}\cline{5-9}
\end{tabular}
\end{table}

In order to increase the overall sensitivity, the three optimized analyses  are then combined, giving rise to seven independent samples. Systematic uncertainties are re-computed for each combination, and a global limit is determined using the modified frequentist approach\cite{cite:freq} including correlations between systematic uncertainties.


\section{Final Results}
The main systematic uncertainties of this search are the background cross-sections~(15\%), the jet energy scale~(from 6 to 17\%), the luminosity calculation~(6.1\%), the effect of parton distribution functions (PDF) uncertainties on signal acceptance~(6\%) and track confirmation~(5\%).

\begin{figure}[t] 
\begin{center}
\begin{minipage}{0.6\linewidth}
\centering
\psfig{figure=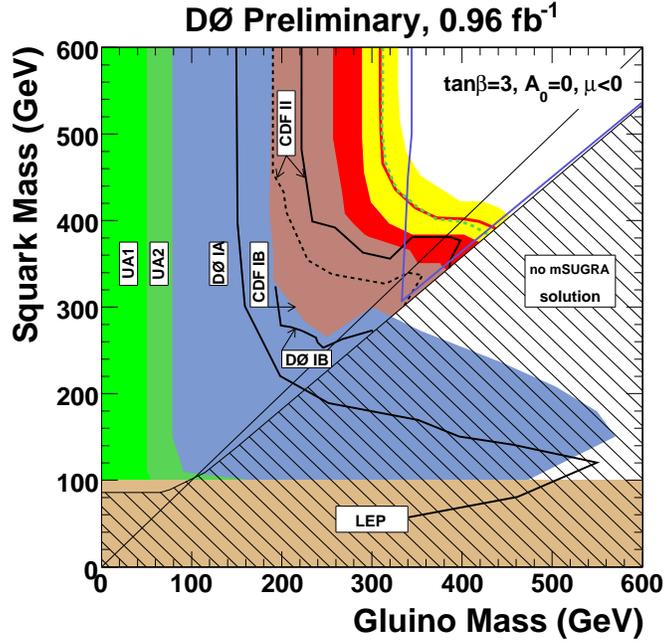,width=\linewidth}
\end{minipage}
\end{center}
\caption{In the squark and gluino mass plane, regions excluded by the analysis at 95\% C.L. in the mSUGRA framework for $tan\beta=3$, $A_0$=0 and $\mu<0$. The red line is the excluded region for the central PDF and renormalization and factorization scale ($\mu_{rf}=Q$). The yellow band is obtained by varying $\mu_{rf}$ by a factor of two and combining with the effects of the PDF uncertainties. 
\label{fig:Exclu1}}
\end{figure}

\begin{figure}[t] 
\begin{center}
\begin{minipage}{0.6\linewidth}
\centering
\psfig{figure=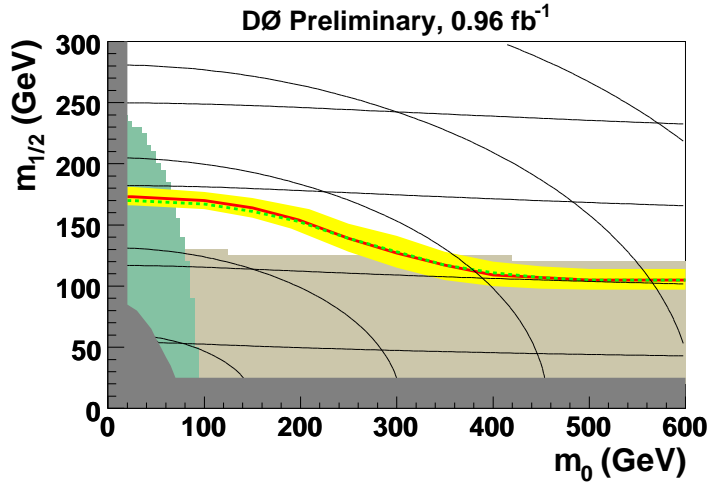,width=\linewidth}
\end{minipage} 
\end{center}
\caption{In the ($m_0$, $m_{1/2}$) plane, regions excluded by the analysis at 95\% C.L. in the mSUGRA framework for $tan\beta=3$, $A_0$=0 and $\mu<0$. The red line is the excluded region for the central PDF and renormalization and factorization scale ($\mu_{rf}=Q$). The yellow band is obtained by varying $\mu_{rf}$ by a factor of two and combining with the effects of the PDF uncertainties. There is no mSUGRA solution in the dark grey region. LEP2 chargino and slepton searches excluded the beige and the green regions respectively. The nearly horizontal thin black lines are the gluino iso-mass curves corresponding to gluino masses of 150, 300, 450 and 600~GeV. The other ones are squark iso-mass curves corresponding to squark masses of 150, 300, 450, 600 and 750~GeV.
\label{fig:Exclu2}}
\end{figure}

The signal cross-section uncertainties are estimated using 41 CTEQ6.1M PDF sets\cite{cite:CTEQ1,cite:CTEQ2}. The PDF effect is combined quadratically with the effect of the renormalization and factorization scale by a factor of two up or down. The three resulting limits are presented in yellow and red color in the squark and gluino mass plane (Fig.~\ref{fig:Exclu1}), and in the ($m_0$, $m_{1/2}$) plane (Fig.~\ref{fig:Exclu2}). The previous limits from Tevatron\cite{cite:D0RunII} and LEP\cite{cite:LEP} are significantly  improved.

Using the most conservative excluded cross-sections, gluino and squark masses are excluded up to 289 and 375~GeV respectively at 95\%~C.L.


\end{document}